\documentclass[12pt, titlepage]{article}
\begin{document}
\title{ Multiple Scales in Small-World Graphs }
\author{ Rajesh Kasturirangan \\
MIT \\ kasturi@ai.mit.edu}
\maketitle
\centerline{\bf \Large Multiple Scales in Small-World Graphs}  \centerline{\large Rajesh %%@
Kasturirangan} \centerline{ \large MIT} \centerline{\large kasturi@ai.mit.edu}
 \footnotesize  \bf 
Small-world architectures may be implicated in a range of phenomena from disease propagation to %%@
networks of neurons in the cerebral cortex [1-2]. While most of the recent attention on small-%%@
world networks has focussed on the effect of introducing disorder/randomness into a regular %%@
graph, we show that that the fundamental mechanism behind the small-world phenomenon is not %%@
disorder/randomness, but the presence of edges of many different length scales. Consequently, in %%@
order to explain the small-world phenomenon, we introduce the concept of multiple scale graphs %%@
and then state the multiple length scale hypothesis. Multiple scale graphs form a unifying %%@
conceptual framework for the study of evolving graphs.  Moreover, small-world behavior in %%@
randomly rewired graphs is a consequence of features common to all multiple scale graphs. To %%@
support the multiple length scale hypothesis, novel graph architectures are introduced that need %%@
not be a result of random rewiring of a regular graph. In each case it is shown that whenever the %%@
graph exhibits small-world behavior, it also has edges of diverse length scales. We also show %%@
that the distribution of the length scales of the new edges is significantly more important than %%@
whether the new edges are long range, medium range or short range. \rm 

\normalsize 
\section{Introduction}A small-world graph [1-2] is a graph obtained by randomly rewiring a small %%@
percentage of the edges of a regular graph. Watts and Strogatz have shown that a relatively small %%@
amount of rewiring dramatically reduces the average path length between points. As a result %%@
small-world graphs remain clustered like regular graphs while having small average path length.
\par
Yet, it is easy to formulate the small-world problem without invoking randomness in any way. %%@
Consider the following definition.\\ \\ 
\bf Definition 1. \rm A graph $N$ is small-world graph if- \\ 
(a) $N$ has small average path length $L_{\tiny \mbox{av}}$ in comparison with a regular graph %%@
with the same number of vertices and edges.\\ 
(b) $N$ is much more clustered than a random graph with the same number of vertices and edges. \\ %%@
\\
Reformulating small-world in this way suggests two questions-
\\ 
(1) How essential is randomness to the small-world phenomenon? In other words, are there other %%@
graph architectures with strong clustering and with low average path length?
\\ 
(2) Is there a principled way of telling when a graph is small-world? Is there a diagnostic %%@
feature of a graph that allows us to do so?
\\\\
To motivate these issues, consider the following three examples:\\\\ 
(A) \bf Social networks \rm: People are unlikely to make new friends at random. Most of the time %%@
new friends are made through old friends. Consider a society where new friends are made by the %%@
"friend of a friend is a friend" rule. How long will it take for such a society to become a %%@
small-world? Note that this is a question about dynamics, i.e., the change in the topology of the %%@
network is not one shot, instead, it takes place over several time steps.
\\ \\
(B) \bf Communication networks \rm : Suppose it is desirable to add a few more connections to a %%@
communications network in order to decrease transmission time. What is the best way to add new %%@
connections?  
\\ \\
(C) \bf Cortical networks \rm : It is quite likely that the cerebral cortex has sub-circuits that %%@
are small-world like. However, the connections in the brain are highly specific. So how do we %%@
know if a given circuit is a "small-world" circuit?
\\ \par
        We show that there is a simple answer to questions (1) and (2), namely, multiple scale %%@
graphs and the multiple length scale hypothesis respectively. The paper is organized as follows- %%@
First, we define multiple scale graphs which is followed by the statement of the multiple length %%@
scale hypothesis. Three different network architectures are introduced. Numerical simulations of %%@
these architectures lend support to the multiple length scale hypothesis. Next, we summarize the %%@
conceptual advantages conferred by multiple scale graphs. Then small-world behavior for randomly %%@
rewired graphs is derived as a consequence of properties common to all multiple scale graphs. %%@
Finally, in the appendix, we explain why multiple scale graphs are inevitable in a small-world %%@
type situation and then illustrate other applications of multiple scale graphs.

\section{Multiple scale graphs and the multiple length scale
hypothesis}
\subsection{Multiple scale graphs}Let $N$ be a graph with $n$ vertices. Let the distance between %%@
two
 vertices, $f$ and $g$ in $N$ be denoted by $\mbox{dist}_N(f, g)$. Let $w$ new edges be added to %%@
$N$, forming a new
 graph $N'$. Call the set of new edges $W$, so that cardinality($W$) = $w$. For each new edge %%@
$e_w$,
 $\mbox{dist}_N (e_w)$ is the distance- $\mbox{dist}_N (f_e, g_e)$ - in $N$  between the %%@
endpoints $f_e$ and $g_e$  of $e_w$. The distribution of length scales in $W$ is measured by the %%@
function-

\begin{equation} D: \ W  \mapsto (1, \infty) \ , \  D(e) = \mbox{dist}_N(e). \end{equation}                                              
\\

\bf Definition 2. \rm  A graph $N'$ obtained by adding edges to the graph $N$ is multiple scale %%@
with respect to $N$ (denoted $N' \preceq N$) if -

\begin{equation} \begin{array}{l} \exists r \gg 0, \ \mbox{ and } \exists \mbox{ length scales } %%@
l_i, i = 1, 2 \ldots  r \mbox{ such that }\\ 0 < l_1 \ll  l_2  \ldots \ll  l_r \ \le n \mbox{ and %%@
}\\ \forall i : i \le r, l_i  \in  D(W).\end{array} \end{equation} 
\\

\bf Definition 3. \rm A graph $N$  is a \bf multiple scale graph \rm if it has a subgraph $M$ %%@
with the same number of vertices such that $N \preceq M$. Note that $M$ contains all the vertices %%@
of $N$ but not all the edges of $N$. \\ \par 

Suppose we start with a regular graph $R_{n,k}$ with $n$ vertices and $k$ edges per vertex and %%@
add $w$ new edges to $R_{n,k}$ forming a new graph $R_{n,k,w}$. Furthermore, if $w \gg %%@
\mbox{log}n$ and the distribution of the new edges is uniform and random-
 
\begin{equation}               R_{n,k,w}  \preceq  R_{n,k} \mbox{  and  }    R_{n,k}  \subseteq %%@
R_{n,k,w} . \end{equation}                              
\\
Therefore $R_{n,k,w}$  is a multiple scale graph. Equation (3) holds  because the new edges are %%@
chosen at random,and as a consequence, cardinality$(D(W))\gg w  \gg \mbox{log}n$.
        Is it possible that the randomly rewired graph $R_{n,k,w}$ is a small-world graph (as %%@
shown in Watts and Strogatz [1]) \it because \rm it is a multiple scale graph? 

\subsection{The Multiple length scale hypothesis} The reduction in average path length when rewiring a graph is %%@
proportional to the number of length scales present in the new edges and the number of edges at %%@
each length scale. In other words, \it any \rm graph, random or deterministic, that is a result of adding sufficiently %%@
many edges at many different length scales to a regular graph, will exhibit small-world behavior. %%@
Moreover, the small-world behavior is not dependent on whether the new edges are long range or %%@
short range. What matters is the \it number of scales \rm represented in the new edges.

\section{Graph architectures}We now review three different graph architectures, the first an analytically tractable %%@
example, the second related to communication networks and the third to social networks. In each case, it is shown that the %%@
numerical data supports the multiple length scale hypothesis.
 
\subsection{ A regular graph coupled with a tree structure (R+T graphs)}  For the sake of simplicity, assume that $n = %%@
k2^m$. Then, we can divide $N$ into $2^m$ blocks, each of size $k$. At this point, edges are %%@
drawn from the midpoint of each block to all the other points in the block, so the R+T graph has %%@
a regular subgraph $R_{n,k}$. Take all the midpoints $m_j$ and number them from $0$ to $2^m -1$. %%@
The tree structure is defined on the set of midpoints of the regular subgraph in the following %%@
manner-  \\            
\begin{equation} \begin{array}{l} \mbox{For each pair of midpoints } m_i \mbox{ and } m_j, \mbox{ %%@
there is a an edge } e_{i,j} \\ \mbox{ connecting } m_i \mbox{ and } m_j \mbox{ if and only if } %%@
i-j \mbox{ is a power of 2, i.e., } \\ \exists \ r \ \le \ m-1 \mbox{ such that } i-j \ = \ 2^r %%@
\end{array}\end{equation} \\ \par 

In an R+T graph, the total number of edges is  $nk/2 + 2^m  \ = \   k^2 2^{m-1} + 2^m $ %%@
which is- $k/2 + 1/k$ - edges per vertex .Therefore, the tree structure adds only $1/k$ %%@
edges per vertex. Moreover, it is not hard to show that the average path length, $L_{\tiny %%@
\mbox{R+T}} \ =  \ m+1 \ \cong \ \mbox{log}_2 m/ k$ . Therefore, in an R+T graph with $n = %%@
256,\  k = 8 \mbox{ and } m= 5$, the number of edges added by the tree structure is 32 and %%@
$L_{\tiny \mbox{R+T}}\   =  \ 6 $. In comparison, in a computer simulation, starting with %%@
the same regular graph, 32 randomly chosen edges were added. We got $L_{\tiny %%@
\mbox{random}}= 7.1  \ >  \ L_{\tiny \mbox{R+T}}$. 
Note that the tree structure forces the R+T graph to be a multiple scale graph, as $R+T %%@
\preceq R_{n,k}$. Moreover, the R+T graph is quite clustered as we are adding only $1/k$ %%@
more edges per vertex.

\subsection{Hub creation and the maximum rule}  Suppose we want to reduce the average path length %%@
in a communications network, where the wiring diagram of the graph is known. In this case, what we may want to do is to %%@
minimize path length while using as few edges as possible. One way of doing this is to create \it %%@
hubs \rm, i.e., vertices that are connected to many other vertices. In the extreme case, we can %%@
create a "universal" hub by adding $n-1$ new edges, from vertex 1 to vertex $ i$ for all vertices %%@
$i$ . If each vertex in the original regular graph has $k \gg \mbox{log} n$ edges then we are %%@
adding only one connection per vertex on average. Moreover, each hub also gives rise to edges of %%@
different length scales. However, this algorithm may not be feasible in general. For example, if this is a model for air %%@
transport, the above graph model leads to overcrowding at the hubs (as has been noticed by everybody who %%@
flies out of a major hub in the U.S).  \par 

Another solution to the rewiring problem in a communications network is the \bf maximum %%@
rule \rm. Let $N_t , \ t = 0,1, 2 \ldots $ be a graph that is evolving in time. At each %%@
time step $t$, a new connection is drawn between the two points that are most distant from %%@
each other in $N_{t-1}$. In the randomized version of the maximum rule, $m$ pairs of points %%@
are selected randomly at each time step t and an edge is drawn between the most distant of %%@
the $m$ pairs (in $N_{t-1}$). The results of a computer simulation of the maximum rule are %%@
displayed in table 1 along with the results for a randomly rewired graph. In the randomly rewired graph, the %%@
rewiring is done assuming a uniform distribution of the new edges. \\ \par
Table 1 shows that for $w$ = 5, 10 and 25, the random rewiring rule does \it better \rm %%@
than the maximum rule which is quite surprising. However, a comparison of $r_{\tiny %%@
\mbox{random}}$ and $r_{\tiny \mbox{max-rule}}$ explains the counterintuitive behavior. For %%@
$w  \ =  \ 5, \ 10 \mbox{ and }25$ we got $r_{\tiny \mbox{max-rule}} \ = \ 3, \ 5 \mbox{ %%@
and } 9$ respectively while $r_{\tiny \mbox{random}}\  = \ 5,\  8 \mbox{ and } 13$ %%@
respectively. Therefore the number of length scales in the randomly rewired graph is more %%@
than in the maximum rule graph if the number of new edges is the same.  \par 
The comparison of the maximum rule and the random rewiring rule suggests that the strongest determinant of small-world %%@
behavior is the presence or absence of a variety of length scales (and not whether the new %%@
edges are short range or long range). This hypothesis was tested  in a computer simulation by selectively %%@
removing length scales from a randomly rewired graph.. The initial graph was a regular graph with $ %%@
\ n \ = \ 275 \mbox{ and } k \ = \ 6$, randomly rewired by adding 50 new edges. Then, %%@
(approximately) half of the edges were removed in three different ways- The shortest 1/2 of %%@
the edges, the middle 1/2 of the edges and the longest 1/2 of the edges. The average %%@
shortest path length was computed for each of the three graphs.The results are shown in Table 2 below.  \par 
 Table 2 shows that for $w \ = \ 27$ there is no difference in the performance of the three %%@
graphs, suggesting that the absolute length of the new edges is not as important as their %%@
distribution. \\ \\ 
\bf Notation. \rm $E^{\tiny \mbox{i}}$ is the number of edges in the initial network %%@
$N^{\tiny \mbox{i}}$. $w$ is the number of new edges. $L^{\tiny \mbox{i}}$ is the average %%@
shortest path length in the initial network $N^{\tiny \mbox{i}}$. $L^{\tiny \mbox{f}}$ is %%@
the average shortest path length in the initial network $N^{\tiny \mbox{f}}$. $r_{\tiny %%@
\mbox{f}}$ is the number of length scales represented by the new edges $w$ with respect to %%@
the distance metric in $N^{\tiny \mbox{i}}$. \\*[.5cm]
\begin{tabular}{|l|l|l|l|l|l|l|l|l|l|l|}\hline \multicolumn{11}{|c|}{\bf Table 1: %%@
Comparison of Maximum rule and Random rewiring rule} \\ \hline & \multicolumn{5}{|c|}{\bf  %%@
Maximum rule } &  \multicolumn{5}{|c|}{\bf Random rewiring rule} \\ \hline $E^{\tiny %%@
\mbox{i}}$ & 825 & 825 & 825 & 825 & 825 &  825 & 825 & 825 & 825 & 825 \\ \hline $w$ & 5 & %%@
10 & 25 & 50 & 100 & 5 & 10 & 25 & 50 & 100 \\ \hline  $r_{\tiny \mbox{f}}$& 3 & 5 & 9 & 15 %%@
& 18 & 5 & 8 & 13 & 19 & 20 \\ \hline $L^{\tiny \mbox{i}}$& 25.14 & 24.63 & 25.00 & 24.25 & %%@
24.83 & 24.81 & 25.16 & 24.90 & 25.42 & 25.14 \\ \hline $L^{\tiny \mbox{f}}$ & 18.05 & %%@
14.12 & 9.60 & 7.12 & 5.42 & 16.42 & 12.73 & 9.97 & 8.11 & 5.97 \\ \hline 
\end{tabular} \\*[1 cm]
\begin{tabular}{|c|l|l|l|} \hline \multicolumn{4}{|c|}{\bf Table 2: Selective removal of %%@
length scales } \\ \hline Length scales removed & 70 - 138 & 4-70 & 30-100 \\ \hline %%@
$n^{\tiny \mbox{i}}$ & 825 & 825 & 825 \\ \hline $w$ & 27 & 27 & 28 \\ \hline $L^{\tiny %%@
\mbox{i}}$ & 24.99 & 24.91 & 24.81 \\ \hline $L^{\tiny \mbox{f}}$ & 10.08 & 10.04 & 10.01 %%@
\\ \hline
\end{tabular}\\

\subsection{ A Friend of a friend is a friend (FOFF)}The FOFF rule is a dynamical rule that is a model of social %%@
graphs . At each time step $t$, an edge is drawn between two vertices $f$ and $g$ having a %%@
common friend $h$ that are not connected by an edge at time $t-1$. Table 3 shows the result %%@
of a computer simulation of the FOFF rule. The decrease in shortest path length for the FOFF rule is much %%@
slower than the corresponding decrease for randomly rewired graphs (see Table 3 below). For $w$  %%@
=  50, 100, 200 and 500 we got $L_{ \tiny \mbox{FOFF}}$ = 18.98, 14.96, 10.23 and 5.48.  %%@
The multiple length scale hypothesis explains the behavior of the FOFF rule quite readily because %%@
$r_{\tiny \mbox{f}}$ grows very slowly as a function of $w$. In fact, for $w$  =  50, 100, %%@
200 and 500 we got $r_{\tiny \mbox{f}}$ = 2, 4, 6 and 19 , which is a lot slower than the %%@
results for the maximum rule or the random rewiring rule. \\*[.5cm]
\begin{tabular}{|c|l|l|l|l|} \hline \multicolumn{5}{|c|}{\bf Table 3: The FOFF rule} \\ %%@
\hline $n^{\tiny \mbox{i}}$ & 825 & 825 & 825 & 825 \\ \hline $w$ &50 & 100 & 200 & 500 \\ %%@
\hline $r_{\tiny \mbox{f}}$ & 2 & 4 & 6 & 19 \\ \hline $L^{\tiny \mbox{i}}$ & 24.95 & 24.76 %%@
& 24.73 & 24.86 \\ \hline $L^{\tiny \mbox{f}}$ &18.98& 14.97 & 10.23 & 5.48 \\ \hline
\end{tabular}\\
\section{A Theoretical discussion of the multiple length scale hypothesis}
Should we study the distribution of length scales in a network or the effects of introducing increasing amounts of disorder %%@
into a regular network? The purpose of this paper is to argue for the former. So far, the multiple length scale hypothesis %%@
has been supported by numerical data. In this section we start with theoretical arguments for the multiple length scale %%@
hypothesis. Then, small-world behavior in a randomly rewired graph $R_{n,k,w}$  is derived %%@
from equation 3 in section II, i.e.,  the fact that $R_{n,k,w}$  is also a multiple scale %%@
graph. 

\subsection{Justifying the multiple length scale hypothesis}
An analysis of the numerical simulations of multiple scale graphs summarized in table 1-3 leads to a broader theoretical %%@
question. \\
(1) What are the conceptual advantages of focussing on multiple-scale graphs? \\ \\ 
\bf Notation. \rm  $N^{\tiny \mbox{i}}$ will always stand for the initial network and $N^{\tiny %%@
\mbox{f}}$ will always stand for the final network. Let the set of new edges be denoted by $W$ %%@
and the cardinality of $W$ = $w$. Throughout the ensuing discussion, $e$ will always stand for a %%@
new edge that has been added to the initial graph $N^{\tiny \mbox{i}}$. For a pair of vertices %%@
$f$ and $g$, the distance between $f$ and $g$ in $N^{\tiny \mbox{i}}$ is denoted by %%@
$\mbox{dist}_{i}(f,g)$ while the distance between $f$ and $g$ in $N^{\tiny \mbox{f}}$ is denoted %%@
by $\mbox{dist}_{f}(f,g)$. Similarly, for each new edge $e$, $\mbox{dist}_{i}(e)$ is the distance %%@
(in $N^{\tiny \mbox{i}}$) between the endpoints $f^e$ and  $g^e$ of $e$. $r$ is the number of %%@
length scales in $N^{\tiny \mbox{f}}$. $D(e)$ is the distibution of  the edges in $W$. Let %%@
$R_{n,k}$ be a regular graph with $n$ vertices, $k$ edges per vertex. Let $R_{n,k,w}$ be the %%@
graph that we get by adding $w$ new edges to $R_{n,k}$ .
\\ \\
   To answer question 1, we present five reasons for studying multiple scale graphs- 
\subsubsection{Generality} In all the numerical simulations, we assumed that the initial network %%@
$N^{\tiny \mbox{i}}$ is a regular network. However, there is no intrinsic reason to do so. The %%@
definition of multiple scale graphs in section II did not assume the existence of an initial graph that is regular. %%@
Similarly, the multiple length scale hypothesis can be phrased as follows: \\ \par 
	Let $N^{\tiny \mbox{i}}$ be a graph with distance metric $\mbox{dist}_{i}(f,g)$. Let %%@
$N^{\tiny \mbox{i}}$ be rewired by adding $w$ new edges to create $N^{\tiny \mbox{f}}$ . Then the %%@
average shortest path length in $N^{\tiny \mbox{f}}$ is proportional to the number of length %%@
scales, $r$, represented by the new edges $e^w$. Therefore, if  $r$ is large,  $N^{\tiny %%@
\mbox{f}}$  has much smaller average path length than $N^{\tiny \mbox{i}}$. In other words, %%@
independent of  the nature of the initial graph $N^{\tiny \mbox{i}}$, distributing new edges at %%@
many length scales is an optimal strategy. \\ \par 
The generality of the multiple scale formulation is particularly useful when we study the dynamics of graphical evolution, where %%@
even if the initial network $N$ is regular, after a few time steps, the network  $N_t$ is not going to be regular. However, at %%@
each stage, $ N_t \preceq N_{t-1} $, so the multiple scale formalism will still be capable of %%@
modelling the dynamics of the graph.
\subsubsection{ Symmetry} Let $N^1$ and $N^2$ be two graphs on the same set of $n$ vertices. Let %%@
$N^2 \setminus N^1$ be the graph with $n$ vertices whose edges are the edges that belong to $N^2$ %%@
but not to $N^1$. Let $N^1 \setminus N^2$ be the graph with $n$ vertices whose edges are the %%@
edges that belong to $N^1$ but not to $N^2$. Then $N^2 \setminus N^1  \ \preceq  \ N^1 \setminus %%@
N^2$ if and only if $N^1 \setminus N^2  \ \preceq \  N^2 \setminus N^1$, i.e., the property of %%@
having multiple scales is \it symmetric \rm (see VII.2.1). 

\subsubsection{Stability} Let  $N^{\tiny \mbox{i},\delta}$  be a perturbation of $N^{\tiny %%@
\mbox{i}}$. If $N^{\tiny \mbox{f}}  \preceq  N^{\tiny \mbox{i}}$, then $N^{\tiny \mbox{f}}  %%@
\preceq  N^{\tiny \mbox{i}, \delta}$ as long as the number of length scales, $r^d$ (of  $N^{\tiny %%@
\mbox{f}}$  with respect to $N^{\tiny \mbox{i},\delta}$ ) satisfies $r^d \sim r$. Consequently, %%@
being multiple scale is a \it stable \rm property of a graph. 

\subsubsection{Ease of diagnosis} Given a graph and its wiring diagram, how can we determine if %%@
the graph is a small-world graph? This is a question that we will have to answer before modelling real-world phenomena %%@
using graphs of any type. Nature rarely tells us the rule that generated a particular graph. It is hard to determine if a graph has %%@
been generated by a random rule or by a deterministic rule. So how can we determine the behavior of the path length of a given %%@
graph? The answer to this question turns on the following observation- \\ 

			 Assume that $N$ is a highly clustered graph with $n$ vertices. Let $R_{n,k}$ be a regular %%@
subgraph of $N$. As a consequence of 4.1.3 , the choice of $k$ does not affect the %%@
distribution of length scales as long as $k \ll n$.  Therefore, if $N  \preceq R_{n,k}$  then  %%@
$N \preceq R_{n,k'}$  as long as $k' \sim k$ . As a result, the number of length scales , %%@
$r_k$ , is independent of  $k$. \begin{equation} \end{equation} 

 Observation (5) above implies that any regular subgraph $R_{n,k}$  can serve as an initial %%@
model. One particularly natural choice of $R_{n,k}$ is the following- Since $N$  has a regular %%@
subgraph, it has a Hamiltonian cycle $H$  that is also a regular subgraph of $N$.  Then, we can %%@
test if $N \preceq H$ or not. If $N \preceq H$, then $N$ is small-world. Furthermore, (O) %%@
guarantees that the choice of $H$ does not affect the outcome. 

\subsection{Deriving small-world behavior from the multiple length scale hypothesis} 
	\bf Tight covering \rm : A length scale $l$ covers the graph $R_{n,k,g}$ if  for each vertex %%@
$h$ in $R_{n,k,g}$, there is a new edge $e, \ \mbox{dist}_{\tiny  \mbox{reg}}(e) \ \cong \ l$  %%@
such that $x$ lies between the endpoints $f$ and $g$ in the regular graph $R_{n,k}$. In other %%@
words, the new edges of scale $l$ wrap around the graph. Furthermore, in order to reduce %%@
average path length, it is not enough that a length scale $l$ cover the graph. Suppose there %%@
are two vertices $f$ and $g$ such that $ \mbox{dist}_{\tiny  \mbox{reg}}(f,g) \cong   l$. %%@
Ideally, we want the shortest path between $f$ and $g$ in $R_{n,k,w}$ to contain edges of %%@
length  $ \cong l$ only. In order to ensure this property of the shortest path, we have to %%@
make sure that the end point of an edge of length  $\cong l$ is close to the starting point of %%@
another edge of length $\cong l$, i.e., the new edges of length $\cong l$ are tightly packed. %%@
In such a situation, we say that a covering of the graph by edges of the length scale $l$ is %%@
\it tight \rm.

Let $R_{n,k,g}$ be a regular graph that has been rewired by adding $w$ new edges. Furthermore %%@
assume that there is a scaling factor $s$ and length scales $s^i, \ i \le \mbox{log}_s n$ such %%@
the new edges are chosen randomly and that their distribution is uniform with respect to the %%@
length scales, i.e.,\\

\begin{equation}
\begin{array}{l}
P \left( e \in W \ : s^k \le \mbox{dist}_{\tiny  \mbox{reg}}(e)  \le  s^{k+1} \right) \  = \ P %%@
\left( e \in W \ : s^l \le \mbox{dist}_{\tiny  \mbox{reg}}(e)  \le  s^{l+1} \right)  \\ \forall \  %%@
0 \le \  k,l \ \le \mbox{ log}_s n 
\end{array}
\end{equation}
\\
Let $k  \cong \  \mbox{ log}_s n $
 and let  $w \cong  \ n \ \ll \ nk$. (6) implies that the number of edges per length scale is %%@
approximately $\frac{n}{\mbox{log}_s n}$. 
If  $s^i   >  \mbox{ log}_s n $
, the edges of that length scale almost certainly cover $R_{n,k,w}$ (see the beginning of this %%@
section for a definition of "cover"). The average distance between successive edges of length  $ %%@
\ s^i$ is about $\mbox{ log}_s n 
 \ \cong \  k$, which means that in $R_{n,k,w}$ successive edges are separated by about an %%@
edge since $ k \cong \ \mbox{log}_s n$. Therefore, if $ s^i > \mbox{ log}_s n$, the edges %%@
of scale $s^i$  tightly cover the graph. 
Let $f$ and $g$ be two vertices such that \\ \\ $ \mbox{           log}_s n \ 
  \le \   s^i  \ \le \  \mbox{dist}_{\tiny  \mbox{reg}}(f,g) \ \le \  s^{i+1},\  i  \le \  \mbox{ %%@
log}_s n $
.\\ \\ The following inequality is an easy consequence of the fact that the scales tightly cover %%@
the graph- \\
\begin{equation}
	\mbox{dist}_{\tiny f}(f,g) \ \le \   2s  + \mbox{dist}_{\tiny f}(h,g)
\end{equation}						
\\
Where $h$ is an intermediate point such that $s^{i-1} \ \le \mbox{dist}_{\tiny  %%@
\mbox{reg}}(h,g) \ \le \ s^i$.  Inequality (7) gives rise to
\\
\begin{equation} \mbox{dist}_{\tiny f}(f,g)   \ \le \   2s. \mbox{log}_s %%@
(\mbox{dist}_{\tiny  \mbox{reg}}(f,g) )
\end{equation}
\\	
showing that the distance in the final graph is uniformly logarithmic for all pairs f and g, %%@
replicating the qualitative behavior of small-world graphs. \\

\bf Remark \rm : The derivation of (8) used the tight covering of $R_{n,k,w}$ at many different %%@
scales, and not on any properties that are unique to random graphs. So (8) holds for all multiple %%@
scale graphs that tightly cover the graph at many different scales. For example, the R+T structure introduced in %%@
section III.1 also satisfies the tight covering property and therefore should behave exactly the same way. 

\section{Conclusion} 
	There is no universal graph architecture that is an optimal solution to all problems. Nevertheless it would be extremely %%@
useful to have a class of graph architectures that are near optimal for a wide variety of problems. Graphs with edges of multiple %%@
sales are an especially rich class of models that provide a unified framework for describing and modeling a variety of graphs. %%@
Multiple scale graphs can be used to model both random and deterministic graphs. This paper shows that properties common to %%@
all multiple scale graphs explains the small-world phenomenon and that the shortest path problem is also much simpler to solve in %%@
multiple scale graphs (see 7.3). Furthermore, the presence of multiple scales in a graph is %%@
easily verified, which makes it an effective tool when applied to real world graphs. \\
Multiple scale graphs are tractable, both analytically and by computer simulations. We will be better equipped to understand the %%@
behavior of real world networks if we focus our attention on the distribution of length scales in graphs as opposed to the effect of %%@
introducing increasing amounts of disorder into a regular graph. Furthermore, every discipline that encounters evolving graph-like %%@
structures will find the theory of multiple scale graphs to be a deep organizing principle.

\section{Methods}
\bf A note about the algorithm and the tables \rm: The same graph algorithm is used in all the %%@
simulations. All the rewiring rules mentioned in this paper are initialized on a regular graph %%@
with $n = 275$ vertices and $k = 6$ edges per vertex. 1000 pairs of vertices are then selected at %%@
random and the shortest distance between all the pairs is found. The algorithm also keeps track %%@
of the number of edges added at any stage.
	All the graphs are plots of the existence or non-existence of a length scale in the new %%@
edges. For a graph with 275 vertices, the value of the function $\mbox{dist}_{\tiny  %%@
\mbox{reg}}(f,g)$ is divided into 20 blocks. For each new edge with vertices $f \mbox{ and } g, %%@
\mbox{dist}_{\tiny  \mbox{reg}}(f, g)$ is computed and then the Y- value of the appropriate block %%@
between 1 and 20 is changed from 0 to 1.

\section{Appendix}Sections 7.1.1-7.1.3 of the appendix summarize the reasons  for the appearance of multiple scale graphs %%@
in clustered small average path length situations. 7.2.1 provides an example of the symmetry property of multiple scale graphs. 7.3 %%@
provides the proof that in multiple scale graphs, not only is the average path length small, the algorithm that computes the shortest %%@
path is also very efficient. 
\subsection{A discussion of multiple scale graphs} In this section, we gather a few remarks that %%@
explain why multiple scale graphs appear whenever there is a competition between shortening %%@
average path length and keeping the number of new edges as small as possible.
\subsubsection{Coarse graining} Assume that the number of new edges, $w$, satisfies $w \ \ll \ %%@
nk$. Let $l$ be a given length scale and $f$ and $g$ be two vertices such that %%@
$\mbox{dist}_{\tiny  \mbox{reg}}(f,g)\ \cong \ l$.   Then, the shortest path between $f$ and $g$ %%@
in $R_{n,k,w}$ is unlikely to contain an edge $e$ with $\mbox{dist}_{\tiny  \mbox{reg}}(e) \ > \ %%@
l$, i.e., the shortest path between two vertices $f$ and $g$ contains new edges that are of %%@
lesser or equal length. The following example helps clarify this point. \\ \\
If $N$ is a regular graph with $n = 2m$ vertices and $k = 2$, i.e.,  $N$ is a ring. If each %%@
vertex is connected to the diametrically opposite vertex, a new graph $N'$ is formed that %%@
is also a regular graph with the same number of vertices,$ n$, and $k = 3$. Therefore the %%@
average shortest distance between vertices is still $O(n)$. The reason for the $O(n)$ %%@
behavior of the shoetest distance is that vertices that are less than $n/4$ away in $N$ are %%@
still the same distance from each other in $N'$, because the new edges are too coarse to be of %%@
any help. \\ \\
	The above example shows that even in the extreme eventuality of all possible maximum length edges being made, the %%@
average distance in the final graph is linear in $n$ and not logarithmic (as one would expect in %%@
a small-world graph). Adding more long range edges does not help reduce the distance between points that are not too %%@
far from one another.

\subsubsection{Redundancy}  Let $f$ and $g$ be two vertices in $R_{n,k,w}$. Then the shortest %%@
path from $f$ to $g$ is unlikely to contain two new edges $e,\  e'$ such that  %%@
$\mbox{dist}_{\tiny  \mbox{reg}}(e) \ > \  ½ \mbox{dist}_{\tiny  \mbox{reg}}(f,g)$and  %%@
$\mbox{dist}_{\tiny  \mbox{reg}}(e') > ½ \mbox{dist}_{\tiny  \mbox{reg}}(f,g)$. As a result, %%@
there is an optimal number of new edges at each scale, and all other new edges are redundant when it %%@
comes to reducing path length.

\subsubsection{ Covering cost} In order to tightly cover the graph with new edges of length  $ \ %%@
\cong  \ l$, we need a minimum of $n/l$ edges. If the total number of new edges is small, larger %%@
length scales are more advantageous. Note that coarse graining and tight covering are opposing demands on the new %%@
edges.  

\subsection{An example of the symmetry of  the relation  $\preceq$}  Suppose the initial graph %%@
$N^{\tiny \mbox{i}}$ has n vertices. Let the edge matrix of $N^{\tiny \mbox{i}}$ be given by the %%@
rule- \\ \\
Rule 1: Two vertices $f$ and $g$ are connected if they differ by a power of 2. \\
Suppose that $N^{\tiny \mbox{i}}$ is rewired by adding edges according to the rule- \\ \\ 
Rule 2: Two vertices $f$ and $g$ are connected if they differ by a power of 3. \\
It is easy to see that rule 2 introduces edges that are of all length scales with respect to rule %%@
1 and vice versa. 

\subsection{ Shortest path algorithms} Given two vertices in a graph how can we find the shortest %%@
path between the two? In a random graph there seems to be no way out besides checking all paths and then picking the %%@
shortest of the lot. From a computational point of view this is an extremely expensive algorithm, %%@
involving $O(n^2)$ computations on average. The same problem seems to arise in the case of a %%@
regular graph modified by adding a few random edges. Even though the graph is small-world and the %%@
behavior of the average shortest path length is known, there is no efficient way of determining the shortest path between two %%@
points. Once again, the multiple length scale hypothesis helps in formulating a better algorithm for the problem of finding the %%@
shortest path between two vertices. \\
Suppose a graph $N$ has many different length scales $s^i, \ i \ \le \ \mbox{log}_s n$  . %%@
Furthermore assume that all the length scales $s^i $ tightly cover the graph $N$.
		Let $f$ and $g$ be two vertices in $N$. Let $h$ be an intermediate vertex in the shortest %%@
path P($f$,$g$) from $f$ to $g$. Define
\begin{equation} \begin{array}{l} k  = max \left\{  l : s^l  \ \le \  %%@
\mbox{dist}_{\tiny  \mbox{reg}}(f,g)\right\} \\
\Delta   =  \left\{  e \ \in  \ W \ : \ s^k  \ \le \  \mbox{dist}_{\tiny  %%@
\mbox{reg}}(f,g) \ \le \   s^{k+1} \right\}\\				       
\Theta \   = \ \left\{ \exists z \ : \  e \ \in \  \Delta \mbox{ such that } e \mbox{ %%@
is an edge from } h \mbox{ to } z \right\} \end{array} \end{equation}		
Then, there is an edge edge $e_h$ in P($f$,$g$) starting at $h$ (going towards $g$) satisfying %%@
the following properties-
	\begin{equation} \begin{array}{l}	e_h  \ \in \  \Delta   \mbox{ and if } y \mbox{ is the %%@
other vertex of } e_h, \mbox{ then } \\
		\mbox{dist}_{\tiny  \mbox{reg}} (y,g)  \ \le \  \mbox{dist}_{\tiny  \mbox{reg}}(y',g)  %%@
\  \ \ \forall \ y' \ \in \ \Theta \end{array}\end{equation}

Two properties of the shortest path algorithm follow from (10) – \\
(i) The shortest path algorithm is local, i.e., at each intermediate vertex $h$ in the shortest %%@
path P($f$,$g$) from $f$ to $g$, the next edge in P($f$,$g$) is in the neighborhood of $h$. \\
(ii)At each intermediate vertex $h$, the shortest path algorithm considers a small subset of the edges in the local neighborhood of %%@
$h$, namely the ones that are of the right scale. 

From (i) and (ii) it follows that the shortest path algorithm performs %%@
$O(\mbox{log}_s(\mbox{dist}_{\tiny  \mbox{reg}}(f,g)))$ computations, which is $ \ \cong  %%@
O(\mbox{log}_s n)$  computations when computing the shortest path from $f$ to $g$. In %%@
comparison, for an arbitrary graph we need to perform $O(n^2)$ computations.\\ \\ \\

\centerline{\bf Acknowledgements}   I want to thank Whitman Richards for many stimulating %%@
discussions about the topic of the paper. \\

 \centerline{\bf References }  

[1] Watts, D. and Strogatz, S., The collective dynamics of small-world networks. \it Nature \rm, Volume 393, 4 June %%@
1998.\\*[.2cm] 
  \par   [2] Watts, D., \it Small Worlds : The Dynamics of Networks between Order and Randomness %%@
\rm , To be published by Princeton University Press.

\end{document}